\documentclass[
 reprint,
 amsmath,amssymb,
 aps,
]{revtex4-2}
\usepackage[english]{babel}
\usepackage{graphicx}
\usepackage{epstopdf}
\usepackage{sidecap}
\usepackage{epstopdf}
\usepackage{ulem}
\usepackage{comment}
\usepackage{dcolumn}
\usepackage{bm}
\usepackage{amsmath}
\usepackage{amsfonts}
\usepackage{amssymb}
\usepackage{latexsym}
\usepackage{color}
\usepackage[dvipsnames]{xcolor}
\usepackage[colorlinks=true, allcolors=blue]{hyperref}
\usepackage{comment}

\newcommand{\charles}[1]{\textcolor{black}{#1}}

\begin{document}

\title{Weibel-mediated filamentary structures observed in the ICF context}

\author{C. Ruyer$^{1,2}$}\email{charles.ruyer@cea.fr}
\author{S. Bola{\~n}os$^{3}$}
\author{P. E. Masson Laborde$^{1,2}$}
\author{L. Gremillet$^{1,2}$}
\author{N. Blanchot$^{1}$}
\author{G. Boutoux$^{1,2}$}
\author{W. Cayzac$^{1}$}
\author{C. Courtois$^{1}$}
\author{S. G. Dannhoff$^{4}$}
\author{V. Denis$^{1}$}
\author{L. Le Deroff$^{1}$}
\author{C. K. Li$^{4}$}
\author{J. Fuchs$^{5}$}
\author{A. Grisollet$^{1}$}
\author{I. Lantuéjoul$^{1}$}
\author{R. Riquier$^{1}$}
\author{R. Smets$^{6}$}
\author{G. D. Sutcliffe$^{7}$}
\author{B. Vauzour$^{1}$}
\affiliation{$^{1}$CEA, DAM, DIF, F-91297 Arpajon, France}
\affiliation{$^{2}$Université Paris-Saclay, CEA, Laboratoire Matière en Conditions Extrêmes, F-91680 Bruyères-le-Châtel, France}
\affiliation{$^{3}$Center for Energy Research, University of California - San Diego, La Jolla, CA, 92093, USA}
\affiliation{$^{4}$Massachusetts Institute of Technology, Cambridge, MA, USA}
\affiliation{$^{5}$LULI - CNRS, CEA, UPMC Univ Paris 06: Sorbonne Université, Ecole Polytechnique, Institut Polytechnique de Paris - F-91128 Palaiseau Cedex, France}
\affiliation{$^{6}$LPP, Sorbonne Université, CNRS, Ecole Polytechnique, F-91128 Palaiseau, France}
\affiliation{$^{7}$Lawrence Livermore National Laboratory, Livermore, CA, USA}

\begin{abstract}
In light of novel and past experimental results, we demonstrate how Weibel-mediated filamentary structures can develop in the expanding plasma plume of a laser-irradiated foil. The transverse ballistic cooling that occurs during the quasi-spherical plasma expansion naturally drives an electron pressure anisotropy, resulting in the growth of electron current filaments. This effect competes with electron-ion Coulomb collisions which tend to isotropize the electron distribution function. Based on theoretical and particle-in-cell modeling, we provide estimates of the dominant wavelength and amplitude of the self-generated magnetic fluctuations, which are found to explain experimental data obtained at the OMEGA and Laser Megajoule facilities.  
\end{abstract}

\maketitle

\section{Introduction}
\label{sec:intro}

The Weibel instability \cite{Weibel_1959, POF_Fried_1959} is a kinetic phenomenon that arises in mostly collisionless plasmas when a pressure anisotropy is present. By amplifying seed magnetic fields and/or current fluctuations, it leads to the segregation of particles according to their current polarity and the formation of current filaments surrounded by magnetic field loops. Understanding the linear and nonlinear properties of the Weibel instability has been the focus of extensive theoretical and numerical work over the past decades \cite{Moiseev_1963, Davidson_1972, Lee_1973, Califano_1998, Kato_2005, Achterberg_2007a, *Achterbeg_2007b, Kaang_2009, Bret_Gremillet_2010, Istomin_2011, Pokhotelov_2011, PRE_Grassi_2017}.

This instability has been widely studied in the context of astrophysical plasmas, where the source of anisotropy is typically provided by counter-propagating plasma flows \cite{POF_Fried_1959, Moiseev_1963, Medvedev_1999, APJ_Drake_2012, Fiuza_2020, PRR_Grassi_2021, Bresci_2022}. In addition, it has attracted significant attention in relativistic laser-plasma interactions, particularly with regard to ion acceleration applications \cite{PRL_Fuchs_2003, PRL_Chen_2009, PRL_Roth_2013, Scott_2017}. It is then triggered by the laser-accelerated electrons, either within the laser irradiation region \cite{Sentoku_2003}, in the collisional target bulk \cite{POP_Gremillet_2002} or over longer spatiotemporal scales far from the laser impact \cite{PRL_Quinn_2012, Ruyer_2020,SA_Zhao_2024}. Its potential development has also been identified in expanding plasmas \cite{Thaury_2010, PRL_Quinn_2012}, though the precise conditions and mechanisms governing its occurrence remain unclear.

In this study, we analyze three experiments performed at large-scale laser facilities that reveal Weibel-type filamentary structures in expanding plasma plumes relevant to inertial confinement fusion (ICF). The nanosecond and millimeter scales of these experiments pose significant challenges for kinetic instability analysis, and self-consistent numerical simulations remain computationally infeasible with current or near-future supercomputers. To overcome these difficulties, we employ a semi-analytical model, developed in Ref.~\cite{True_85}, which explains how weak electron pressure anisotropy is spontaneously driven during spherical plasma expansion and makes way for the growth of the Weibel instability.

Our paper is organized as follows. In Sec.~\ref{sec:th}, we review the basic ingredients of the model, and explain how it can be applied to predict the development of electron anisotropy in plasma plumes and the properties of the resulting Weibel magnetic filaments. We then perform particle-in-cell (PIC) kinetic simulations \cite{Lefebvre_2003} to validate our analytical estimates. In Sec.~\ref{sec:xp}, we compare our findings with three independent experimental measurements, confirming our understanding of the underlying physics and the ability of our model to predict, at least qualitatively, the onset conditions for the Weibel instability in expanding laser plasmas. Finally, in Sec.~\ref{sec:ccl}, we discuss the implications of our results to ICF experiments.

\section{The Weibel instability in an expanding plasma}
\label{sec:th}

\subsection{Spherical plasma expansion as a source of electron thermal anisotropy}
\label{sec:model}

We start by describing the physical mechanisms underlying the development of electron thermal anisotropy during the expansion of a plasma sphere. We explain how the conservation of electron angular momentum leads to a decrease in transverse electron pressure, which in turn drives the Weibel instability.

\begin{figure}
    \includegraphics[width=0.35\textwidth]{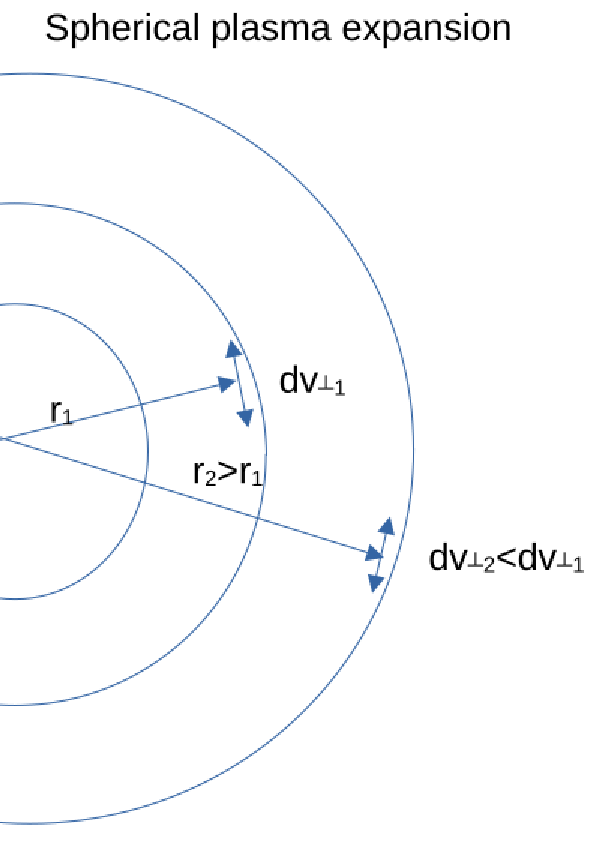}
    \caption{ \label{fig:sketch}
    Sketch of a spherical expansion. Because of ballistic effects, particles that reach position $r_1$ have more transverse thermal velocity spread than the particles that reach position $r_2>r_1$, thus driving a pressure anisotropy. Electron-ion Coulomb collisions balance this mechanism.}
\end{figure}

During the expansion of a hot (quasi-) spherical plasma, the fast and light electrons drag the heavier ions via space-charge fields. In a collisionless medium, the assumption of a central force field constrains the radial evolution of the electron distribution. Specifically, conservation of the electron angular momentum $l\equiv m_e v_\perp r$ ($m_e$ is the electron mass and $v_\perp$ its transverse velocity) causes the transverse electron temperature (or mean transverse kinetic energy) to decrease as $T_\perp \propto r^ {-2}$. If the radial temperature drops slower than $r^{-2}$, a momentum anisotropy naturally arises, with the hotter direction aligned along the radial axis. This mechanism,  illustrated in Fig.~\ref{fig:sketch}, is counteracted by electron-ion Coulomb collisions, which tend to isotropize the electron distribution. Therefore, the development of the Weibel instability, which feeds on the electron momentum anisotropy, hinges on the competition between the anisotropy growth resulting from the expansion and its mitigation due to collisions.

The kinetic model we adopt, originally proposed by True~\cite{True_85}, is based on the following key assumptions regarding a spherically expanding plasma. First, in the central and dense part of the plasma, the Coulomb collisions are assumed to dominate, enforcing an isotropic Maxwellian electron distribution. Beyond a critical radius $r_\star$, where the electron density is $n_e(r=r_\star) = n_e^\star$, the plasma is considered sufficiently collisionless to allow a slight electron pressure anisotropy to arise. The value of $n_e^\star$ is related to the electron mean free path, $\lambda_\mathrm{mfp}$, through $\int_{r_\star}^\infty dx/ \lambda_\mathrm{mfp}(x)=1$, yielding
\begin{equation}\label{eq:nes}
    n_e^\star \simeq \frac{3\times 10^{23}\, [{\rm cm^{-3}}]}{Z \ln \Lambda} \, \frac{(3T_e [{\rm keV}]/2)^2}{L[{\rm \mu m}]} \,,
\end{equation}
where $Z$ is the ion charge, $\ln \Lambda$ is the Coulomb logarithm, and $L$ is the plasma density scale length. We have neglected electron-electron Coulomb collisions, an approximation valid for $Z \gg 1$.

For a collisionless, initially Maxwellian plasma, the conservation of total electron energy, $E = \frac{1}{2}m_e v^2 - e\phi(r)$ ($e$ is the elementary charge) and electron angular momentum, $l = m_e v_\perp r$, is employed to parameterize the electron distribution function in the form~\cite{True_85}:
\begin{align}\label{eq:fe}
    f_e &= F_0 n_e^\star\exp\left( - \frac{ m_ev_r^2-2e\phi}{2T_e} - \frac{m_ev_\perp r - l_\star }{l_0} \right) \nonumber \\
    &\quad {\rm if}\quad m_ev_\perp r>l_\star \,, \nonumber \\
    f_e &= F_0 n_e^\star\exp\left( - \frac{ m_ev_r^2-2e\phi}{2T_e} \right) \nonumber \\
    &\quad {\rm elsewhere.}
\end{align}
The normalization factor is $F_0 = (m_e T_e)^{-3/2} (2\pi)^{-1/2}$.
This model distribution captures two key regimes: for $r < r_\star$, the plasma retains its Maxwellian character, while for $r > r_\star$, particles with large transverse momenta $m_e v_\perp r$ are increasingly restricted from accessing larger radii, as illustrated in Fig.~\ref{fig:sketch}.

\begin{figure*}
    \includegraphics[width=0.3\textwidth]{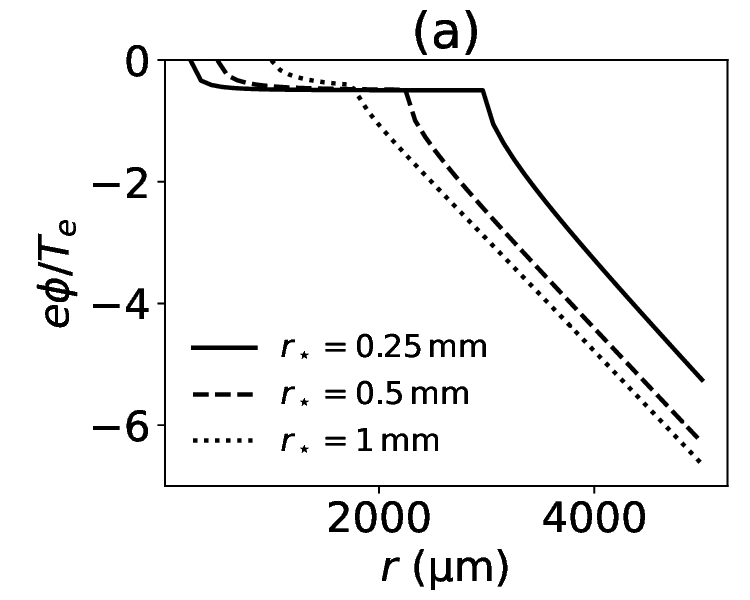}
    \includegraphics[width=0.3\textwidth]{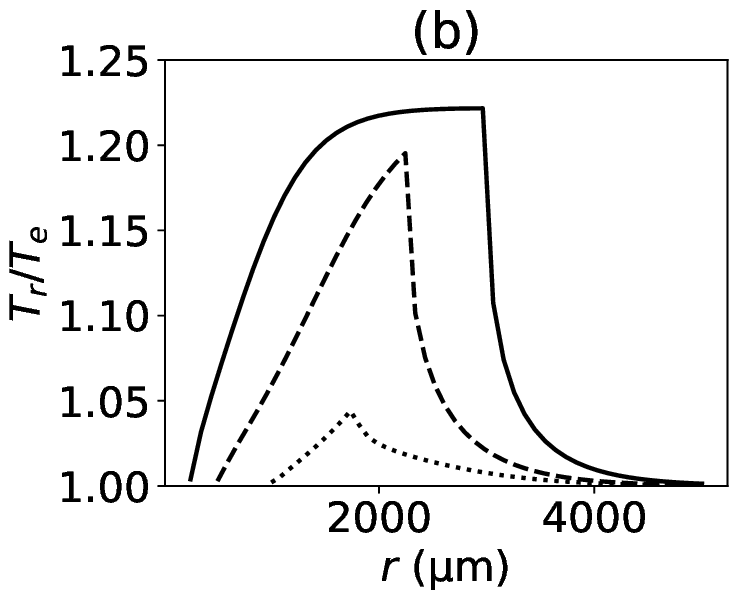}
    \includegraphics[width=0.3\textwidth]{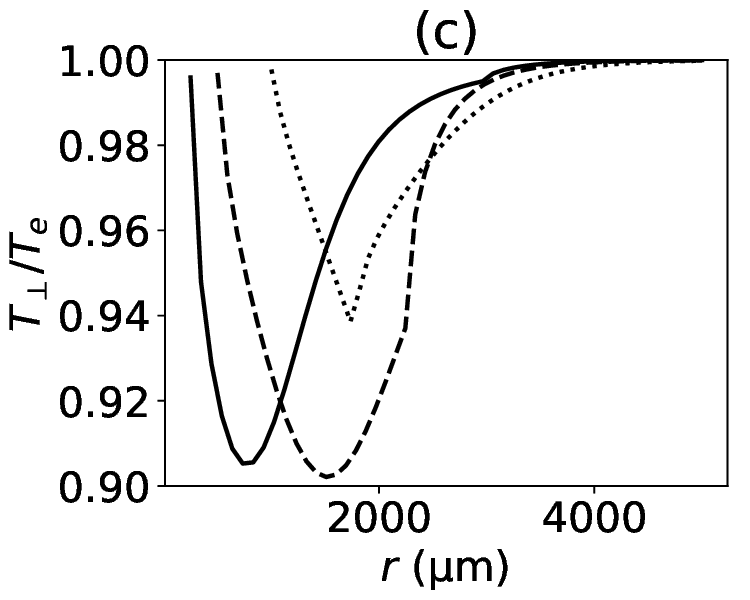}
    \caption{\label{fig:th}
    Model results for the expansion of a carbon plasma ($Z=6$, $A=12$) with density profile $n_e = n_e^\star \exp[-(r-r_\star)/L]$. The plasma is characterized by $L= 0.5\,\rm mm$, $T_e=3\,\rm keV$ and $\ln\Lambda = 10$, leading to $n_e^\star \simeq 2\times 10^{20}\,\rm cm^{-3}$  based on Eq.~\eqref{eq:nes}. Shown are the spatial profiles of (a) the electrostatic potential, (b) the radial temperature and (c) the transverse temperature for $r_\star = 0.4\,\rm mm$ (solid), $0.5\,\rm mm$ (dashed) and $1\,\rm mm$ (dotted).
    }
\end{figure*}

The boundary angular momentum $l_\star$ is derived by applying conservation of energy and angular momentum to a particle originating from $r = r_\star$. This yields the radial velocity:
\begin{equation}
    v_r(r) = \sqrt{\frac{2}{m_e} \left( E + e\phi(r) - \frac{l^2}{2m_e r^2} \right)} \,.
\end{equation}
Without loss of generality, we set $\phi(r=r_\star)=0$, so that $\phi(r>r_\star)$ is negative to confine the electrons in the collisionless region. For $v_r(r_\star)$ to remain real and positive, the condition $l^2 < 2m_e r_\star^2 [E + e\phi(r_\star)]$ must hold. This constraint defines the boundary angular momentum as 
\begin{equation}\label{eq:ls}
    l_\star^2 = 2m_e r_\star^2 E
    \,.
\end{equation}

The angular momentum $l_0$ in Eq.~\eqref{eq:fe} determines the transition from the region where the electron distribution function remains isotropic to the region where the Coulomb collisions are weak enough to allow a finite pressure anisotropy. 
It is evaluated in Ref.~\cite{True_85} from the Fokker-Planck equation, but a simpler derivation can be obtained from the variance of the electron-ion Coulomb collision scattering angle \cite{Jackson_chap13, Sentoku_2008},
\begin{equation}
    \delta \theta^2/4 \simeq \nu_{ei} \delta t\,,
\end{equation}
valid for small angles. The scattering angle $\delta \theta$ is related to $l_0$ and the typical electron angular momentum, $r_\star \sqrt{m_e T_e}$, through $\theta \simeq l_0/(r_\star \sqrt{m_eT_e})$. The time interval $\delta t$ over which the deflections occur is governed by hydrodynamics, $\delta t \simeq L/c_s$, where $c_s=\sqrt{Z_iT _e/m_i}$ is the sound speed. Only electron-ion collisions are considered here. While the relevant time scales justify discarding ion-ion collisions, neglecting electron-electron collisions, valid for $Z\gg 1$, is primarily for simplicity. Using $\nu_{ei} \simeq \sqrt{T_e/m_e}/\lambda_{\rm mfp}$ leads, up to a numerical factor of the order of unity, to
\begin{equation} \label{eq:l0}
    l_0\simeq r_\star \left(\frac{Lm_eT_e}{3\lambda_{\rm mfp}} \right)^{1/2}\, \left(\frac{m_i}{Zm_e}\right)^{1/4} \,,
\end{equation}
which matches the expression given in Ref.~\cite{True_85}.

For a quasineutral plasma expansion, the electron number density is obtained by integrating Eq.~\eqref{eq:fe} over $E$ and $l$ (see Appendix \ref{app:ne}):
\begin{align}
    n_e(r) &= n_e^\star \left[ e^{\frac{e\phi(r)}{T_e}} - \left(1-\frac{r^2}{r_\star^2 }\right)^{1/2} e^{\frac{e\phi(r)}{T_e (1-r^2/r_\star^2)}} \right] \nonumber \\
    &+ \bar{n_e}(\phi,r)\,, \label{eq:phi}
\end{align}
where $\bar{n}_e(\phi,r)$ is given by  Eq.~\eqref{eq:nebar_final} in the Appendix.

Given an electron density profile $n_e(r)$, a boundary radius $r_\star$ and an unperturbed electron temperature $T_e$ (for $r<r_\star$), we use Eq.~\eqref{eq:phi} to derive the electrostatic potential $\phi(r)$ upon which the electron distribution function $f_e$ [Eq.~\eqref{eq:fe}] depends. We can then evaluate the temperature anisotropy
\begin{equation}
    a = \frac{T_r}{T_\perp} -1
    \label{eq:a} 
\end{equation}
from the electron radial ($T_r$) and transverse ($T_\perp$) temperatures
\begin{align}
    T_r=&\frac{\int dv_r dv_\perp v_\perp v_r^2 f_e(v_r,v_\perp)  }{\int dv_r dv_\perp v_\perp f_e(v_r,v_\perp) } \,, \label{eq:tr}\\
    T_\perp=&\frac{\int dv_r dv_\perp v_\perp^3 f_e(v_r,v_\perp)/2 }{\int dv_r dv_\perp v_\perp f_e(v_r,v_\perp) } \,. \label{eq:tp}
\end{align}
This model is valid provided $T_r$ and $T_\perp$ remain close to $T_e$.

Figure~\ref{fig:th} illustrates the model predictions for a fully ionized carbon plasma (charge state $Z=6$ and atomic mass $A=12$) with electron temperature $T_e=3\,\rm keV$. We consider a density profile of the form $n_e = n_e^\star \exp[-(r-r_\star)/L ]$ with scale length $L= 0.5\,\rm mm$, leading to $n_e^\star = 2 \times 10^{20}\,\rm cm^{-3}$ according to Eq.~\eqref{eq:nes} for $\ln \Lambda = 10$. We first numerically extract $\phi(r>r_\star)$ from Eq.~\eqref{eq:phi} as shown in Fig.~\ref{fig:th}(a) for three values of $r_\star = 0.4\,\rm mm$ (solid lines), $0.5\,\rm mm$ (dashed lines) and $1\,\rm mm$ (dotted lines). We then proceed to evaluate $T_r$ and $T_\perp$ using Eqs.~\eqref{eq:fe}, \eqref{eq:phi}, \eqref{eq:tr} and \eqref{eq:tp}.

One can see that the radial [Fig.~\ref{fig:th}(b)] and transverse [Fig.~\ref{fig:th}(c)] temperatures exhibit opposite behaviors: heating for $T_r$ and cooling for $T_\perp$. Both effects are modest ($\lesssim 5\,\%$), intensify as the collisional radius $r_\star$ decreases, and occur primarily at finite distances  ($r \simeq 1$--$2\,\rm mm$).

Notably, for $r_\star = 0.4\,\rm mm$, the electric potential exhibits a millimeter-long plateau. Consequently, conservation of the effective electron energy, $T_r/2+T_\perp-e\phi$, dictates that the energy lost in the transverse directions (the cooling effect) is transferred predominantly to the radial direction (the heating effect). As the critical radius $r_\star$ increases--corresponding to a more planar expansion--this plateau vanishes, limiting radial heating. Overall, smaller collisional radii $r_\star$ result in larger thermal anisotropies. The final level of anisotropy is a trade-off between transverse cooling (which causes $a$ to increase) and the isotropizing effect of Coulomb collisions.

Quasi-spherical plasma expansions, as modeled here, are common in direct-drive ICF experiments. In this context, several works have reported filamentary structures in the plasma corona~\cite{Rygg_2008, Seguin_2012, Manuel_2013}. The authors of Ref.~\cite{Manuel_2013} ascribed their observations to the magneto-thermal instability \cite{Tidman_1974}, which has a growth rate of $\sim 10 \,\rm ns^{-1}$. They argued against the Weibel instability because the plasma is too collisional in the filamentary region. Indeed, the hydrodynamic density profiles used in Ref.~\cite{Manuel_2013} indicates  $n_\star\simeq 8.6\times 10^{19}\,\rm cm^{-3}$ (Eq.~\eqref{eq:nes} for $L\simeq 400\,\rm \mu m$ and $T_e=1.1\,\rm keV$), leading to $r_\star \simeq2\,\rm mm $. Hence, the filamentary structures observed in that experiment reside within the collisional and isotropic region $r \lesssim  1.5\, {\rm mm} < r_\star$.

The mechanisms driving electron pressure anisotropy in spherical expansions can also operate, to some extent, in simple exploding foils, if the plasma is viewed far enough from the dense target. In this configuration, multiple beams can be focused onto the same region, producing a hotter corona than in spherical irradiation with equivalent total laser power. In this case, the characteristic radius $r_\star$ correlates with the on-target beam waist, rather than the target size. Moreover, because electron pressure may remain anisotropic as long as the laser drives the expansion, any triggered Weibel instability should correlate temporally with the laser drive.

All of the experiments analyzed in Sec.~\ref{sec:xp} correspond to the exploded-foil configuration and exhibit Weibel-mediated filamentary structures probed by proton radiography~\cite{RSI_protograhya, RSI_protograhyb}. Since these measurements are sensitive to electromagnetic fields, interpreting them requires predicting the magnetic field amplitude, a task addressed in the next section.

\subsection{Growth rate, wavenumber and saturated amplitude of the Weibel magnetic fields}
\label{sec:weibel}

\begin{figure*}
    \includegraphics[width=0.3\textwidth]{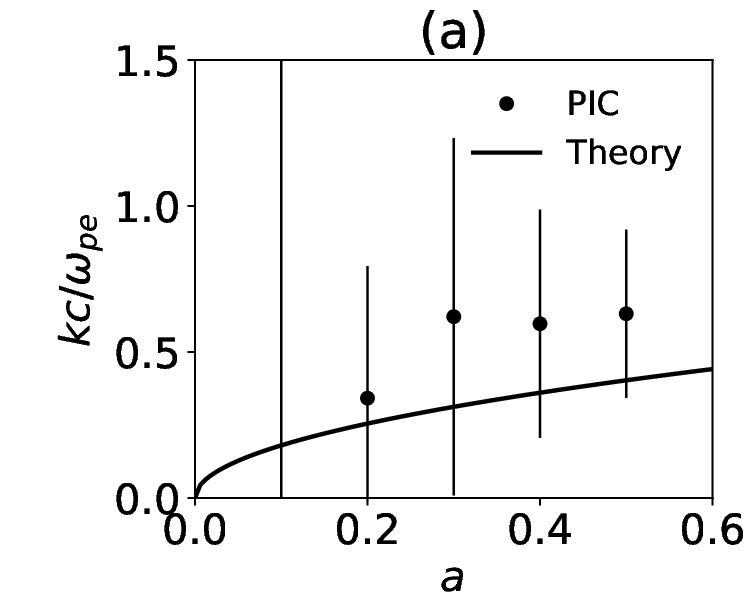}
    \includegraphics[width=0.3\textwidth]{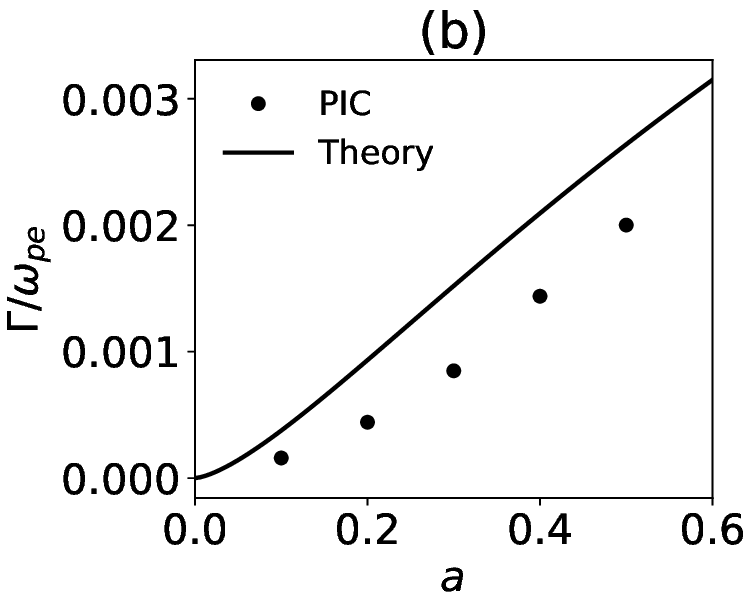}
    \includegraphics[width=0.3\textwidth]{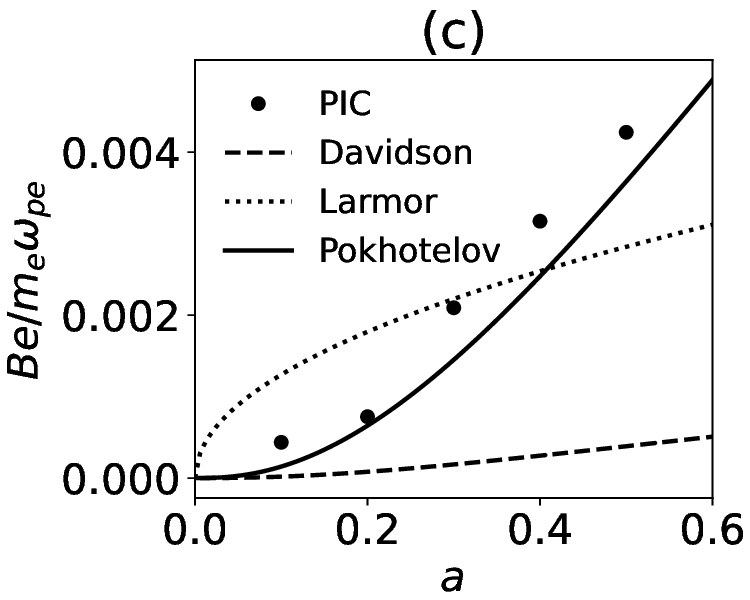}
    \caption{ \label{fig:bnc}
    Dominant magnetic wavenumber (a), growth rate (b) and magnetic field amplitude $B=\langle 2B(x)^2\rangle_x^{1/2}$ (c) from collisionless PIC simulation as markers.
    Expressions \eqref{eq:knc} and \eqref{eq:gnc} are superimposed as solid lines on panels (a) and (b).
    (a) The dominant wavelength (as markers) and spectrum width (as an error bar) are extracted from the simulation results through the two first momentum of $\vert B(k)\vert ^2$ in the $k$-space.
    (c) Estimates of Eqs.~\eqref{eq:dav}, \eqref{eq:lar} and \eqref{eq:pok} are plotted as dashed, dotted and plain lines, respectively. 
}
\end{figure*}

In a collisionless plasma, the maximum growth rate ($\Gamma_\mathrm{W}$) and the corresponding wavenumber ($k_\mathrm{W}$) of the electron Weibel instability are related to the anisotropy ratio ($a$) and the transverse temperature ($T_\perp$) through~\cite{Krall} 
\begin{align}
    \Gamma_\mathrm{W} &\simeq 0.31\, \frac{\omega_{pe}}{c} \sqrt{\frac{T_\perp}{m_e}} \frac{a^{3/2}}{a+1} \,, \label{eq:gnc} \\
    k_\mathrm{W} &\simeq 0.57\, \frac{\omega_{pe}}{c} a^{1/2} \,. \label{eq:knc} 
\end{align} 
Here, $\omega_{pe}$ is the electron plasma frequency and $c$ is the speed of light. \charles{Note that if the instability was driven by the ion population, both the growth rate and dominant wavevector would be proportional to the ion plasma frequency,  thus giving much smaller values for  a similar anisotropy.} These expressions hold if the plasma collisionality is negligible relative to the magnetic field growth--specifically, if the electron mean free path is larger than the magnetic wavelength and the momentum exchange rate is much smaller than $\Gamma_\mathrm{W}$.

To validate these theoretical estimates, we perform one-dimensional (1D), collisionless PIC simulations using the \textsc{calder} code~\cite{Lefebvre_2003}. The periodic spatial domain spans $819\,c/\omega_{pe}$ with a mesh size $\Delta x = 0.1\,c/\omega_{pe}$ and a time step $\Delta t = 0.095\,\omega_{pe}^{-1}$. Third-order interpolation is employed. The homogeneous plasma consists of fully ionized titanium ions ($Z=22, A=47.8$) with an initial temperature $T_i=0.7\,\rm keV$. Electrons are initialized as a two-temperature Maxwellian population with thermal anisotropy $a$. Their temperature transverse to the spatially resolved $x$--axis is set to $T_y=T_z \equiv T_r = 1\,\rm keV$, yielding a longitudinal temperature $T_x \equiv T_\perp = T_r/(a+1)$. The electron and ion species are represented by $10^4$ macroparticles per cell per species.

By increasing the electron thermal anisotropy from $a=0.1$ to 0.5, we obtained different growth rates and dominant wavenumbers. The effective growth rate ($\Gamma_B$) is inferred from the logarithmic derivative of the total magnetic energy ($\Xi_B$) during the exponential phase:  
$\Gamma_B = (1/2) d \ln(\Xi_B)/dt$. The dominant wavenumber $k_B$ and spectral width $\Delta k_B$ are computed as
\begin{align}
    k_B =& \frac{\int k  \vert B(k)\vert^2dk }{\int \vert B(k)\vert^2 dk } \,, \label{eq:kB} \\
    (\Delta k_B)^2 =&\frac{ \int (k-k_B)^2 \vert B(k)\vert^2 dk }{\int \vert B(k)\vert^2dk } \,. \label{eq:DeltakB}
\end{align}

Both quantities are reported in Fig.~\ref{fig:bnc}(a) as markers and error bars, while $\Gamma_B$ is plotted in Fig.~\ref{fig:bnc}(b). Formula~\eqref{eq:knc} [solid lines in Fig.~\ref{fig:bnc}(a)] shows fair agreement with the PIC predicted wavenumbers for $a \ge 0.2$. The associated spectral widths [error bars in Fig.~\ref{fig:bnc}(a)] are significant, though decreasing at larger $a$. The lower-anisotropy, and hence slower-growth, case $a=0.1$ is characterized by a width of $\Delta k \simeq 6\,\omega_{pe}/c$, making the dominant wavenumber ($k_B \simeq 2.5\,\omega_{pe}/c$, not shown) poorly representative. By contrast, formula~\eqref{eq:gnc} captures the simulated growth rates well across all cases [see Fig.~\ref{fig:bnc}(b)].

We now examine three different theoretical estimates of the magnetic field amplitude ($B_{\rm sat}$) at saturation. The latter quantity is extracted from the PIC simulations at the end of the exponential phase via $B_{\rm sat} = \langle 2 B(x)^2 \rangle_x^{1/2}$ (where $\langle \cdot \rangle_x$ denotes the spatial average) and is illustrated in Fig.~\ref{fig:bnc}(c).

The most common model, proposed by Davidson~\cite{Davidson_1972}, assumes that saturation arises from the transverse trapping of particles within the magnetic filaments. A nearly monochromatic magnetic spectrum results in transverse electron oscillations around the $B$-field nodes at the bounce frequency $\omega_\beta = \sqrt{e k v_r B/m_e}$, where $v_r$ is the thermal velocity along the ``hot'' (here radial) direction. An estimate of the saturated $B$-field amplitude is obtained by balancing the Weibel growth rate with the bounce frequency:
\begin{align}
    B_{\rm sat} &\simeq \frac{m_e}{e} \frac{\Gamma_{\rm W}^2}{k_{\rm W}}\sqrt{\frac{m_e}{T_\perp}} \nonumber \\
    &\simeq 0.3\, \frac{m_e}{e} \sqrt{\frac{T_\perp}{m_e}} \frac{a^{2}}{(a+1)^{5/2} } k_{\rm W} \,. \label{eq:dav}
\end{align}
This trapping-based model is widely used in the literature and has proven accurate for Weibel instabilities induced by fast electrons~\cite{PRL_Gode_2017} or ion-driven current filamentation instabilities~\cite{PRL_Swadling_2020}. However, to the best of our knowledge, it has not been tested for weakly unstable systems such as those addressed here. In fact, the broad magnetic spectrum indicated by the error bars in Fig.~\ref{fig:bnc}(a) is scarcely compatible with the assumption of a quasi-monochromatic spectrum required for this model. Consequently, Eq.~\eqref{eq:dav} [dashed line in Fig.~\ref{fig:bnc}(c)] significantly underestimates the saturated $B$-field amplitude observed in our simulations.

Another estimate of $B_{\rm sat}$ can be obtained assuming that the electrons become fully magnetized within the filaments, that is, when their Larmor radius becomes comparable with the filament wavelength \cite{Moiseev_1963, Bresci_2022}. This yields
\begin{equation}
\label{eq:lar}
    B_{\rm sat} \simeq \frac{m_e}{e} \sqrt{\frac{T_e}{m_e}} \frac{k_\mathrm{W}}{2\pi} \,.
\end{equation}
Plotted as a dotted line in Fig.~\ref{fig:bnc}(c), the $a^{1/2}$ scaling of this formula is contradicted by the steeper dependence exhibited by the PIC results.

Finally, the broad spectrum of growing wavenumbers suggests that a quasilinear model may better describe the instability saturation. Within this approach, Pokhotelov~\cite{Pokhotelov_2011} derived the following expression for the saturated field strength:
\begin{equation}
\label{eq:pok}
    B_{\rm sat} \simeq 2.25\,\frac{m_e}{e} \sqrt{\frac{T_e}{m_e}} \frac{a^2}{(a+1)^{5/2}} k_{\rm W} \,.
\end{equation}
Surprisingly, this formula exhibits the same scalings with $a$ and $T_\perp$ as the trapping-based formula above, differing only by a prefactor that is $\sim 7.4\times$ larger. As shown by the solid line in Fig.~\ref{fig:bnc}(c), this formula provides a fairly good match to the PIC results.

In summary, we have evaluated three estimates of the saturated $B$-field strength. While the models based on transverse trapping or full magnetization of the plasma electrons do not align with our numerical data, the quasilinear estimate is found to provide a good match. We therefore recommend using Eqs.~\eqref{eq:gnc}, \eqref{eq:knc}, and \eqref{eq:pok} to predict the magnetic field amplitude in weakly anisotropic, collisionless plasmas.

\section{Comparison with experimental ICF-relevant observations}
\label{sec:xp}

In light of previous theoretical prediction, we now analyze several high-energy laser experiments where Weibel-type filamentary structures have been characterized.

\subsection{Expansion of a hohlraum window at OMEGA}
\label{sec:peml}

\begin{figure*}
    \includegraphics[width=0.95\textwidth]{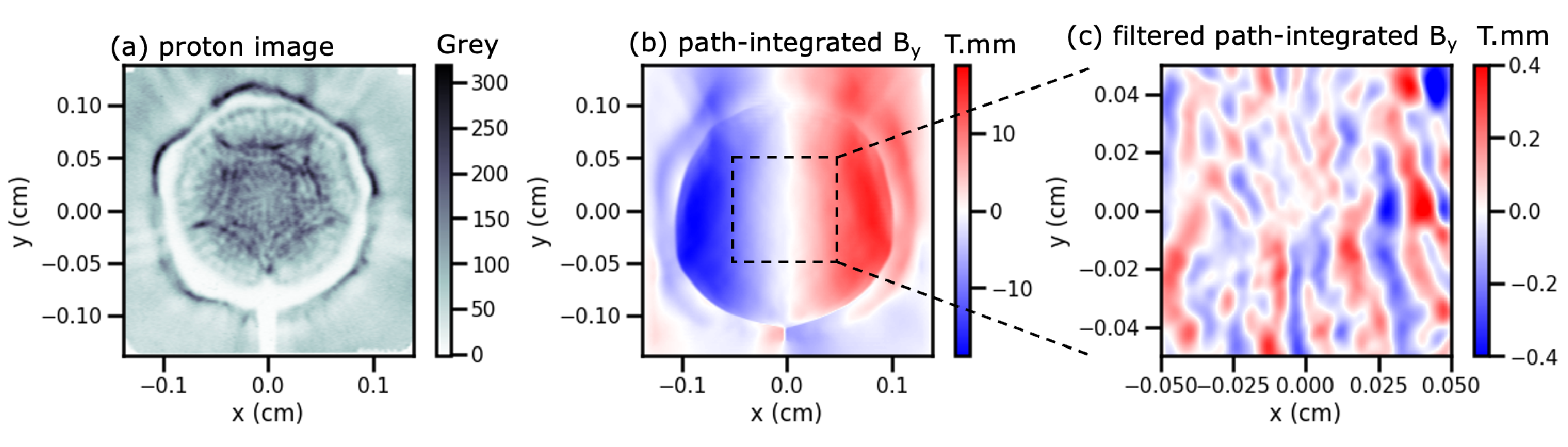}
    \caption{ \label{fig:omega_analysis}
    Proton radiograph obtained at OMEGA (a) and the corresponding reconstructed path-integrated magnetic field (b). Panel (c) shows the filtered path-integrated magnetic field, highlighting the small-scale magnetic structures that produce the dots observed in the proton radiograph.
}
\end{figure*}

\begin{figure}
    \includegraphics[width=0.45\textwidth]{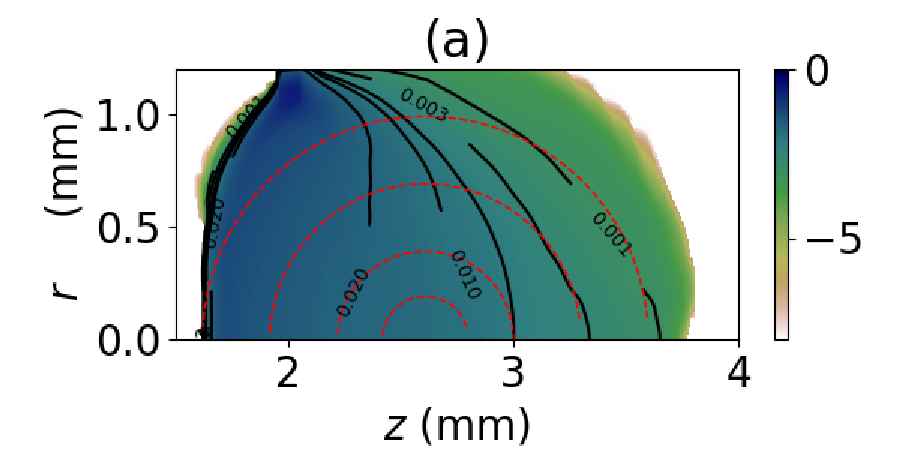} \\
    \includegraphics[width=0.45\textwidth]{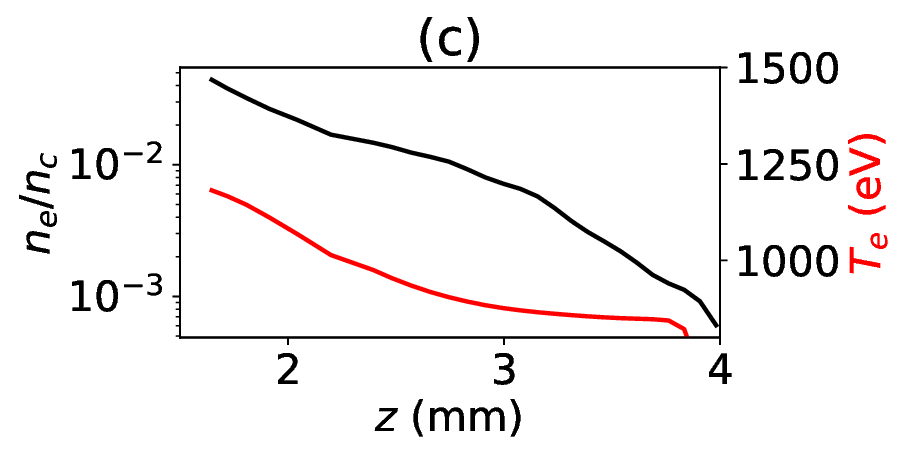} \\
    \includegraphics[width=0.45\textwidth]{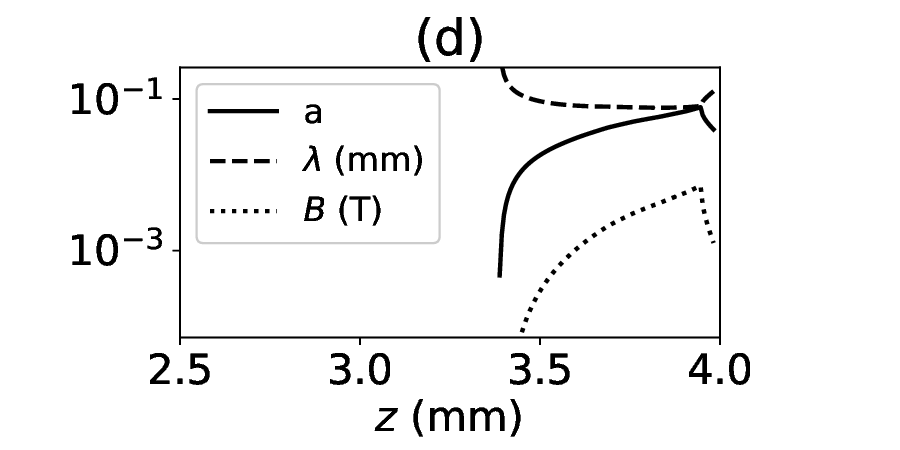}
    \caption{\label{fig:peml}
    Results of the hydro-rad \textsc{troll} simulation of the experiment described in Ref.~\cite{Masson_2019}. (a) Electron plasma density (normalized to the laser critical density $n_c$) of the expanding plastic window in the ($r,z$) plane at $t=1.2\,\rm ns$. Concentric circles of radii $0.4$, $0.6$, $0.9$, and $1.2\,\rm mm$, centered at $z = 2.4\,\rm mm$, are superimposed (red dashed lines) to indicate the transition to spherical expansion. (b) Electron density and temperature profiles along the symmetry axis $r=0$. (c) Predicted electron thermal anisotropy, $B$-field amplitude and wavelength using the density profile of panel (b) for $T_e = 1050\,\rm eV$, $Z=3.5$ and $A=6.5$.
}
\end{figure}

The first experiment analyzed was performed at the OMEGA facility. It involved the irradiation of a gold cylindrical hohlraum, probed axially by a D$^3$He capsule proton backlighter. As described in Ref.~\cite{Masson_2019}, the expansion dynamics and associated fields of the gold bubble driven by ten OMEGA beams ($0.351\,\mu\rm m$ wavelength, $\sim 500\,\rm J$ total energy in a $\sim 1\,\rm ns$ square pulse) could be reproduced numerically in the absence of a gas fill or plastic windows. Here, we examine additional unpublished shots performed with a gas fill and closing plastic foils.

The proton radiographs of the irradiated target, such as the one displayed in Fig.~\ref{fig:omega_analysis}(a), exhibit dots with a typical wavelength of a few hundred microns, superimposed on the larger structures discussed in Ref.~\cite{Masson_2019}. These dots are consistent with current filaments perpendicular to the plastic window that are probed longitudinally [along the $z$ axis in Fig.~\ref{fig:omega_analysis}(a)], similar to the measurements discussed in Ref.~\cite{Ruyer_2020}.

To characterize these filaments (in terms of size and field strength), we infer the path-integrated magnetic field using the \href{https://github.com/flash-center/PROBLEM}{\textsc{problem}} solver \cite{Bott_2017} [Fig.~\ref{fig:omega_analysis}(b)]. We then isolate the $B_y$ field component associated with the dot structures by applying a low-pass filter with a cutoff at $350\,\mu\text{m}$ [Fig.~\ref{fig:omega_analysis}(c)]. This analysis reveals that the magnetic fluctuations of interest are characterized by a wavelength of $\lambda \sim 100$–$200\,\mu \rm m$ and a path-integrated field strength $\int B\,dl \simeq 0.1$--$0.2\,\rm T\,mm$.

We now apply our model to predict the Weibel fields induced by the electron thermal anisotropy emerging within the expanding plastic foils used as windows in the hohlraum. These results are then compared with the experimentally inferred field characteristics. To this end, we conducted a two-dimensional axisymmetric hydro-radsimulation using the \textsc{troll} code~\cite{Lefebvre_2018}, modeling the gas-filled hohlraum as described in Ref.~\cite{Masson_2019}. A closeup of the expanded window [Figs.~\ref{fig:peml}(b,c)] reveals that at the probing time ($t = 1.2\,\rm ns$), the plastic foil spans $2\,\rm mm$. The plasma density profile follows the expected exponential decay with a scale length of $\sim 0.34\,\rm mm$, while the electron temperature reaches $1050\,\rm eV$ for $r \gtrsim 3\,\rm mm$. These parameters indicate low collisionality, which validates the estimates from Sec.~\ref{sec:model}. From Eq.~\eqref{eq:nes}, we derive $n_e^\star \simeq 1.8 \times 10^{-2} n_c$.

In Fig.~\ref{fig:peml}(a), the density isocontours (black curves) are compared to concentric circles of radii $0.4$, $0.6$, $0.9$, and $1.2\,\rm mm$, centered at $z = 2.4\,\rm mm$ (where $n_e(z) = n_e^\star$). At sufficient distances from the foil plane, the overlap of the isocontours and these circles confirms the transition toward spherical expansion. This suggests that the spherical expansion model is applicable near the axis for $r \gtrsim 0.3$--$0.5\,\rm mm$ and $z \gtrsim 3\,\rm mm$ (or equivalently $n_e \lesssim 7 \times 10^{-3} n_c$). Notably, the density isocontours exhibit negative curvature at $r = 0$ and $z \lesssim 2.4\,\rm mm$ due to the geometry of the hohlraum entrance hole.

By extracting the density profile along the $z$-axis for $n_e \leq n_e^\star$ and setting $r_\star \simeq 0.3\,\rm mm$ (with $Z = 3.5$ and $A = 6.5$), we obtain the electron thermal anisotropy [solid line in Fig.~\ref{fig:peml}(c)], which peaks at $a \simeq 0.11$ near $z \gtrsim 2.8\,\rm mm$. Although the model in Ref.~\cite{True_85} strictly requires $Z \gg 1$, which is only marginally satisfied here, its predictions do not change significantly when electron-electron Coulomb collisions are included in the mean-free-path calculation [Eq.~\eqref{eq:l0}], pointing to a sub-dominant effect of self-collisions.

Although the predicted filament wavelengths [$\lambda_{\rm W} \equiv 2\pi/k_{\rm W} \simeq 20$–$100\,\mu \rm m$, dashed line in Fig.~\ref{fig:peml}(c)] are somewhat lower than the experimental measurements ($\lambda \simeq 100$–$200\,\mu\text{m}$), the ranges overlap, indicating reasonable consistency. Note that the finite size of the proton source prevents characterization of filaments with wavelengths below $\sim 50\,\mu \rm m$~\cite{PRL_Li_2006}. \charles{The inverse of the Weibel growth rate is around $6\,\rm ps$ and thus compatible with the nanosecond scale of the plasma expansion. On the contrary, an ion-driven instability however gives a growth time of $\sim 0.3\, \rm ns$, not consistent with the experiment.} The predicted field strength reaches $B_{\rm sat} \simeq 0.06\,\rm T$ in the region $2.5 \leq z \leq 3\,\rm mm$, leading to a path-integrated $B$-field on both target sides at measurable levels ($\sim 0.006\,\rm T\,mm$), albeit a factor of two below the experimental values. Our prediction should therefore be considered a lower bound. Overall, the interplay between spherical expansion and Coulomb collisions yields a weak yet non-negligible anisotropy, capable of driving sub-tesla magnetic filaments throughout the isothermal corona, resulting in detectable proton deflections.

\subsection{Exploding-foil experiment at OMEGA}
\label{sec:sutcliffe}

\begin{figure}
    \includegraphics[width=0.45\textwidth]{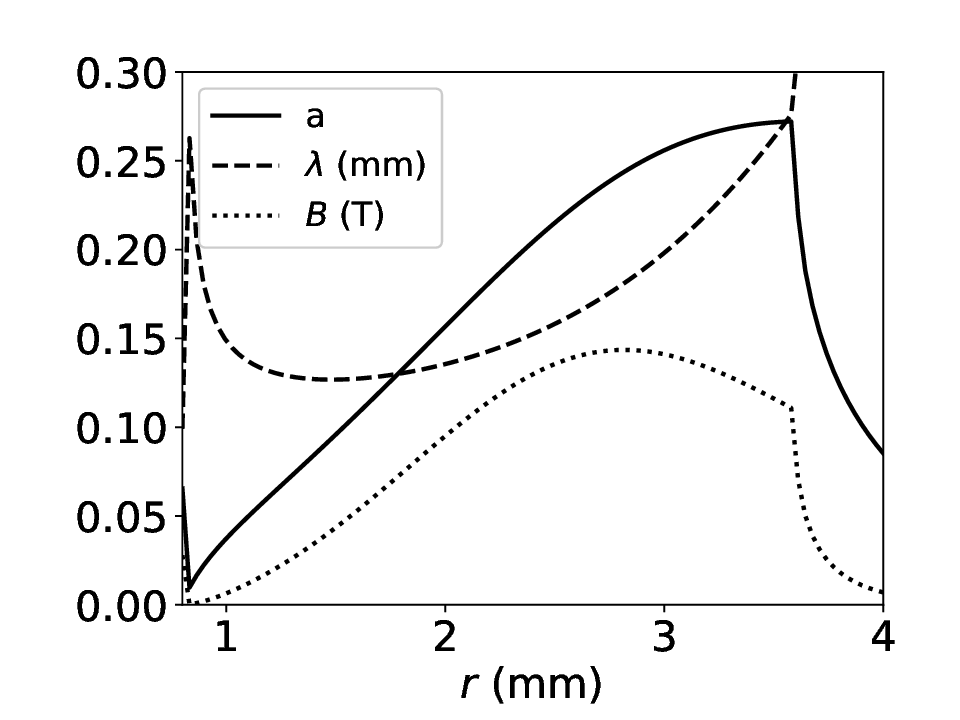}
    \caption{Predicted electron thermal anisotropy (solid line), filament wavelength (dashed line) and $B$-field amplitude (dotted line) from the model of Sec.~\ref{sec:th}, for $Z=3.5$, $A=3.5$, $T_e=0.5\,\rm keV$ and an exponentially decreasing density profile with a scale length of $L = r_\star = 800\,\rm \mu m$. \label{fig:sutcliffe}
}
\end{figure}

Another experiment performed at the OMEGA facility by Sutcliffe \text{et al.}~\cite{Sutcliffe_2022} reports well-characterized filamentary structures in the plasma expanding from a solid foil. In this study, two plastic foils are irradiated with the OMEGA beams. Fast protons generated by a capsule implosion probe one target transversely and the other normally. The proton radiographs obtained during the first nanosecond of the experiment do not exhibit any filaments; however, those taken after $2\,\rm ns$ evidence interesting features developing in the expanding plasma~\cite{Sutcliffe_2022b}. While the face-on view also captures the large-scale Biermann battery fields, the side-on measurements allow for a clean analysis of the filaments extending over $\sim 2\,\rm mm$. These filaments exhibit a path-integrated $B$-field of 0.05--$0.1 \,\rm T\,mm$ and a typical wavelength of $\lambda \sim 200\,\rm\mu m$.

High-quality Thomson measurements give a good idea of the plasma parameters during the foil expansion. They suggest that an electron temperature of $\sim 0.5-0.7\,\rm keV$ is reached after $1.5\,\rm ns$. In the corona, these measurements indicate a plasma scale length of $\sim 750\,\rm \mu m$, comparable with the laser spot diameter. Applying our model with $L=r_\star \simeq 750\, \rm \mu m$, $T_e = 2T_i = 0.5\,\rm keV$, $Z=3.5$ and $A=6.5$ yields an anisotropy profile that extends over $\gtrsim 2\,\rm mm$ and reaches $a \simeq 0.05$ (solid line in Fig.~\ref{fig:sutcliffe}). As in the previous section, the low-$Z$ material used here results in a weakly collisional regime, with an electron mean free path an order of magnitude above the predicted magnetic wavelength. Thus, aside from the neglect of electron-electron Coulomb collisions,  we are confident in the validity of the collisionless estimates of Eqs.~\eqref{eq:gnc} and \eqref{eq:knc}. We therefore expect the dominant wavelength to range between $\lambda_{\rm W} \simeq 120$ and $250\,\rm \mu m$ (dashed line) with a magnetic field amplitude reaching $B_{\rm sat} \simeq 0.02\,\rm T$ (dotted line). This corresponds to a line-of-sight-integrated field of $\sim 0.015\, \rm T\,mm$, consistent in order of magnitude (though somewhat lower) with observations. The spherical expansion mechanism described in Sec.~\ref{sec:model} thus appears capable of driving the electron thermal Weibel instability in nanosecond-timescale,  millimeter-long expanding plasmas, as characterized in Refs.~\cite{Sutcliffe_2022, Sutcliffe_2022b}.
\charles{
On the contrary, the ion-driven instability has a growth rate that is too small for the experiment.}

\subsection{Exploding-foil experiment at LMJ}
\label{sec:lmj}

\subsubsection{Experimental setup and results}

\begin{figure}
    \includegraphics[width=0.5\textwidth]{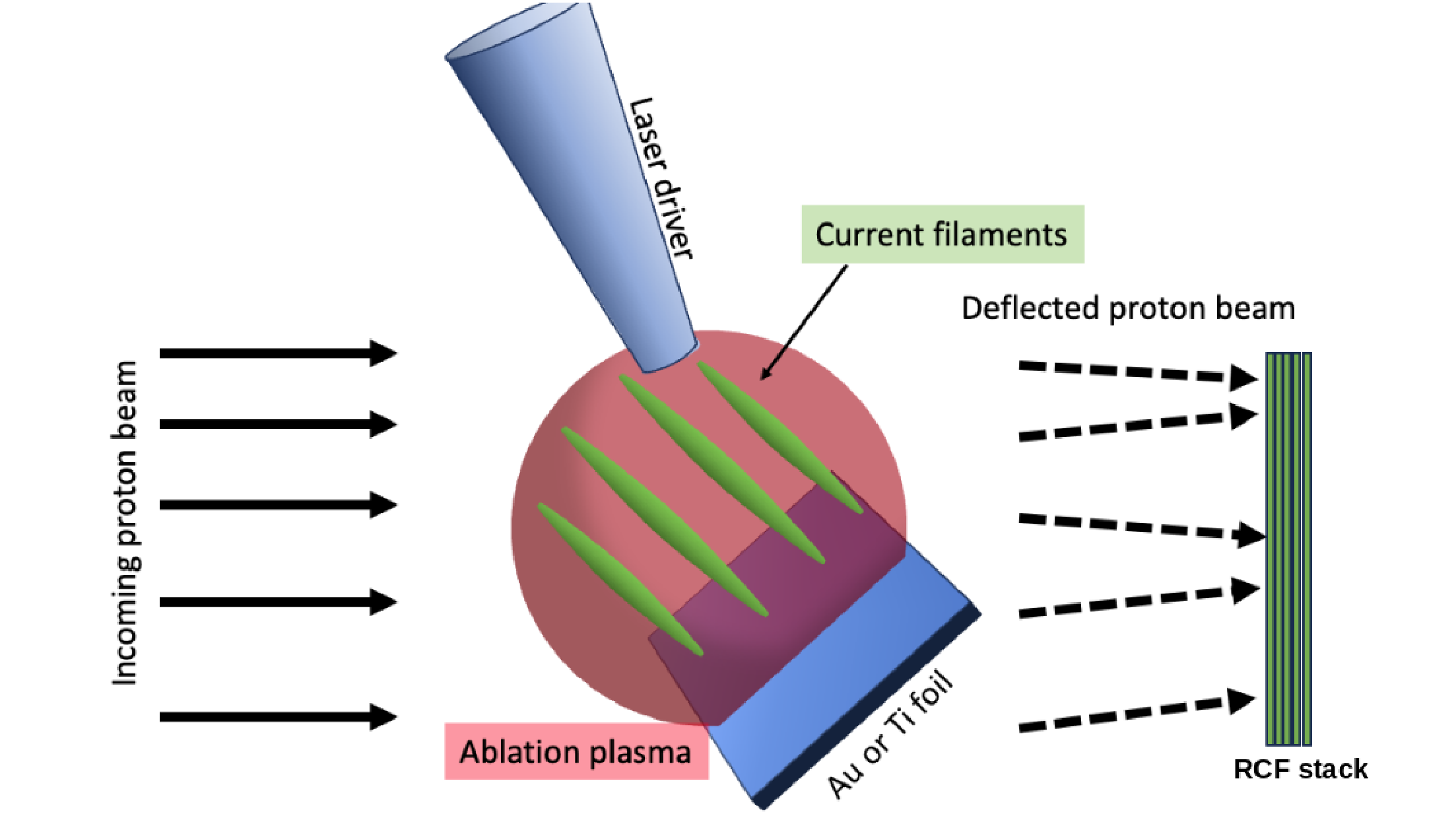}
    \caption{ \label{fig:lmj_setup}
    Experimental setup at the LMJ-PETAL facility. A (Ti or Au) foil target is irradiated by LMJ quads over $4.5\,\rm ns$. Fast protons produced by the short-pulse PETAL beam probe the electromagnetic fields induced during the plasma expansion \cite{HEDP_PETAL}. Representative proton radiographs are shown in Fig.~\ref{fig:lmj_p+}.}
\end{figure}

\begin{figure}
    \includegraphics[width=0.5\textwidth]{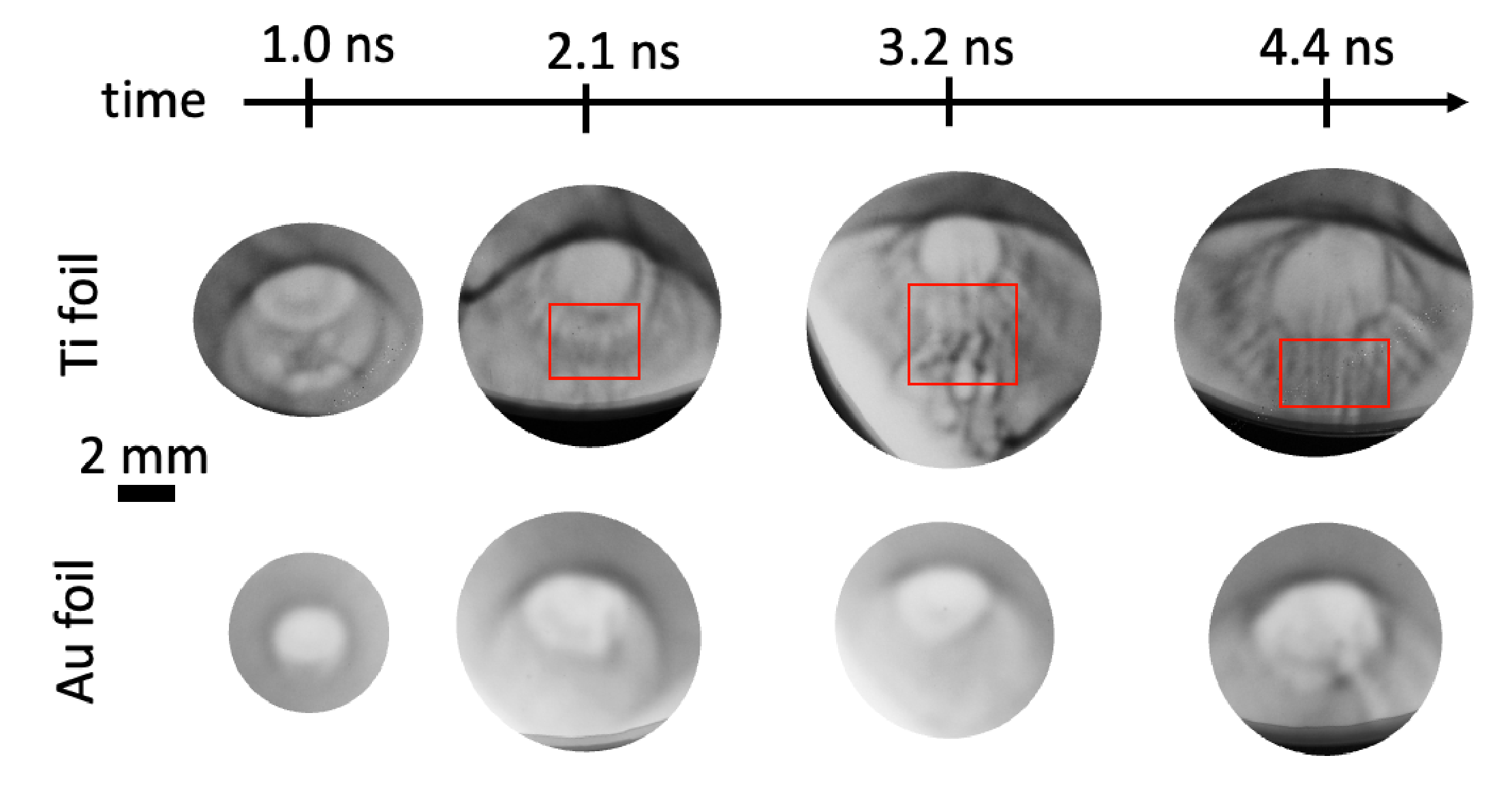}
    \caption{ \label{fig:lmj_p+}
    %
    Proton radiographs (for 29~MeV protons) at four different times. The top and bottom rows show the results for the Ti and Au foils, respectively. Rectangles indicate the region where the filamentary structures are analyzed (see Fig.~\ref{fig:lmj_B}).}
\end{figure}

\begin{figure}
    \includegraphics[width=0.5\textwidth]{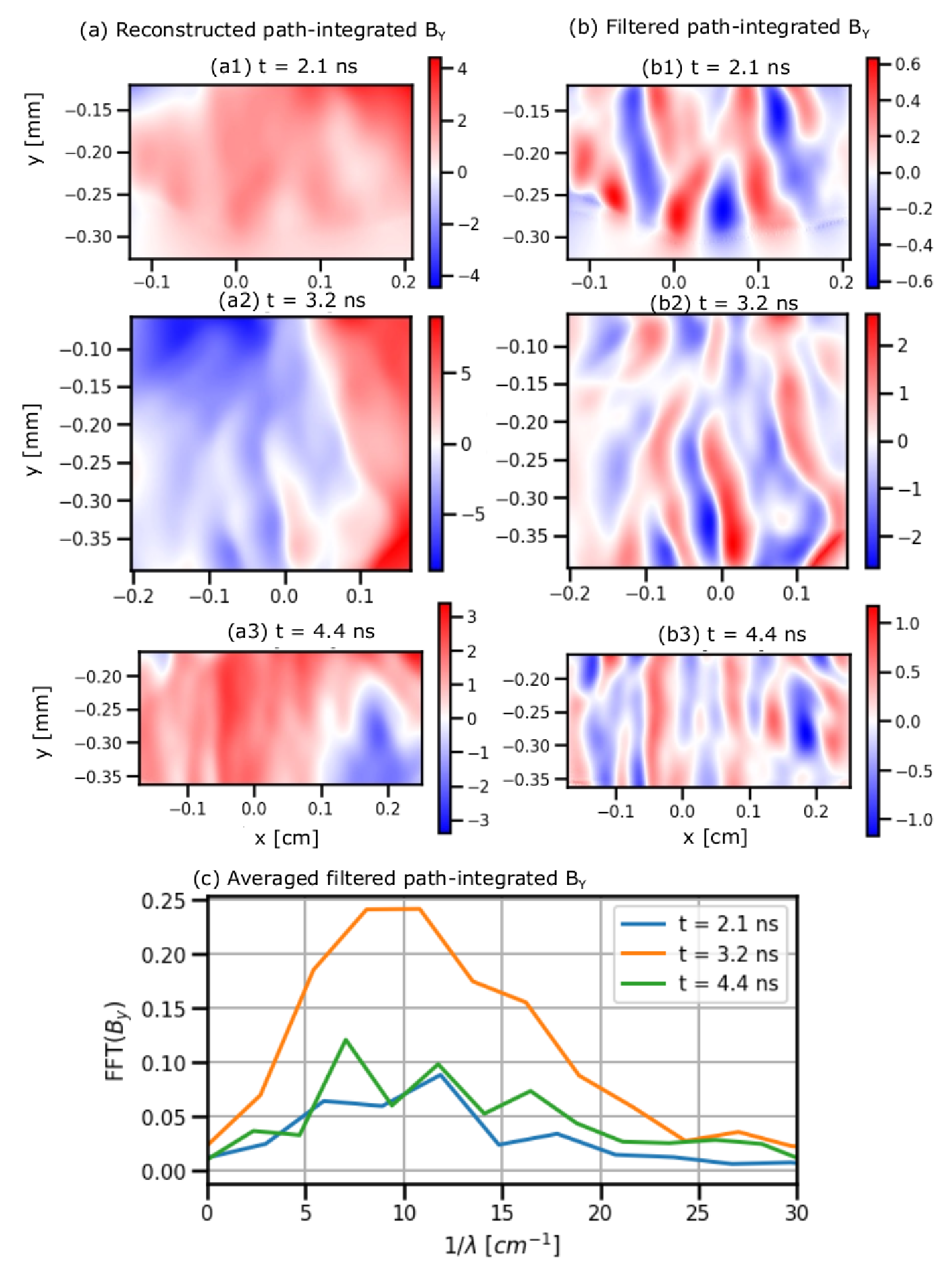}
    \caption{ \label{fig:lmj_B}
    Reconstructed distributions of the path-integrated $B_y$ field ($\int B_y\,dz$ in $\rm T\,mm$ units) at three successive times ($t=2.1$, 3.2, and $4.4\,\rm ns$) from the Ti shot in Fig.~\ref{fig:lmj_p+}. (a1)-(a3) Closeups of the region delineated by the red rectangle in Fig.~\ref{fig:lmj_p+}.(b1)-(b3) Filtered distributions highlighting the filamentary fluctuations. (c) Spatial spectra (along $x$) of the $y$-averaged, filtered path-integrated $B_y$ profiles shown in panels (b1)-b3).}
\end{figure}

The last experiment analyzed was conducted at the LMJ-PETAL facility (see setup in Fig.~\ref{fig:lmj_setup}). The target was irradiated by four LMJ quads focused at different locations on the target surface and with different timings [Fig.~\ref{fig:lmj_setup}]. Each quad delivered $10\,\rm kJ$ of energy over a $4.5\,\rm ns$ square pulse, resulting in an average on-target intensity of $5\,\times 10^{14}\,\rm W/cm^2$. To probe the fields induced in the expanding target plasma, a proton beam was accelerated by the short-pulse ($< 1\,\rm ps$), ultraintense ($>10^{18}\,\rm W/cm^2$) PETAL beam interacting with a $25\,\rm \mu m$ thick gold foil \cite{HEDP_PETAL}.

Figure~\ref{fig:lmj_p+} displays the proton radiographs obtained using a titanium (Ti) (top row) or a gold (Au) foil (bottom row). Note that proton scattering within the solid foils themselves induced beam broadenings of $\sim 77\,\rm \mu m$ in Ti and $\sim 240\,\rm \mu m$ in Au. For the Ti foil, the radiographs reveal the development of filamentary structures after $2.1~\rm ns$. By contrast, the Au case shows smoother proton dose modulations, suggesting much weaker fields, if any. Qualitatively similar filamentary structures have been observed previously on the LMJ at the front side of irradiated aluminum foils \cite{Bott_2021}.

We reconstructed the magnetic field following the procedure described in Sec.~\ref{sec:peml}. Accordingly, the path-integrated $B$-field reached $0.6$, $2.0$, and $1.0\,\rm T\,mm$ after $2.1$, $3.2$ and $4.4\,\rm ns$, respectively. In addition, the distance between filaments was found to increase over time. Specifically, Fourier analysis revealed that the wavelength of the dominant mode was of $0.8~\rm mm$, $1.1~\rm mm $ and $1.5~\rm mm$ at the considered times [Fig.~\ref{fig:lmj_B}(c)].

\begin{figure*}
    \includegraphics[width=0.3\textwidth]{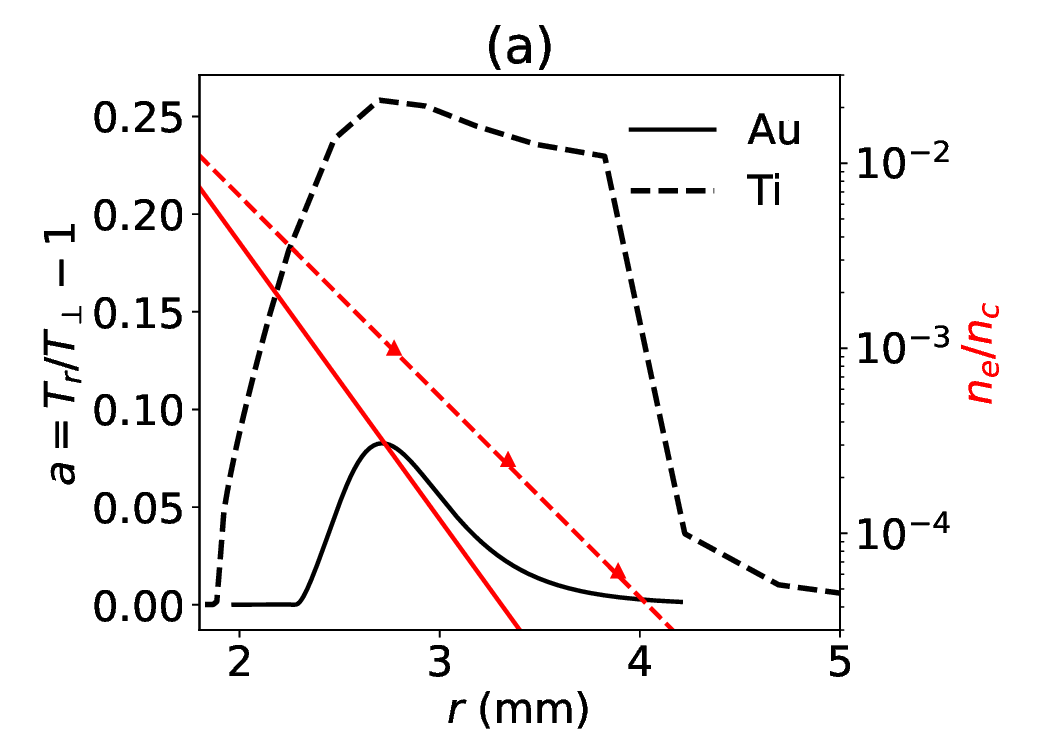}
    \includegraphics[width=0.3\textwidth]{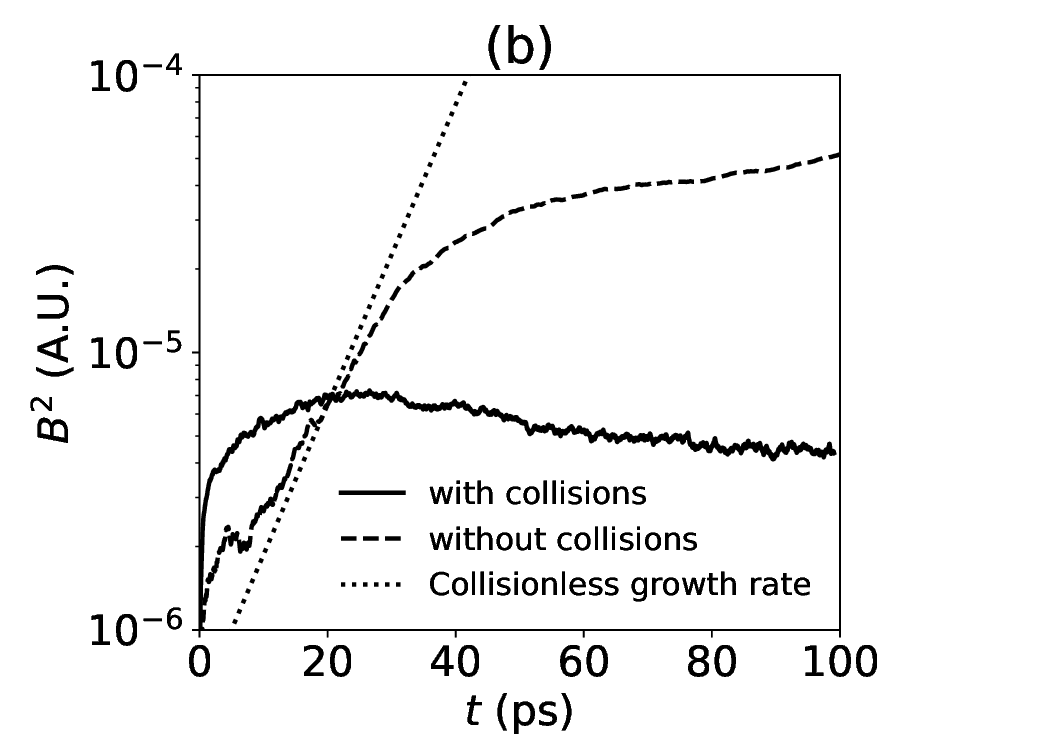}
    \includegraphics[width=0.3\textwidth]{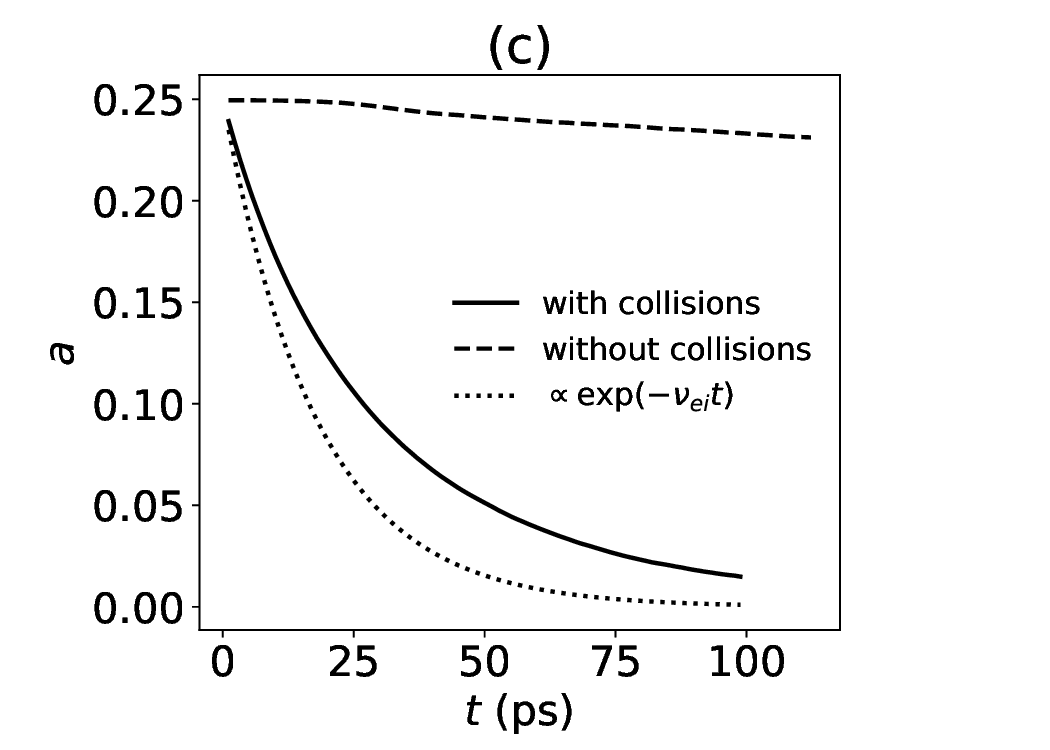} \\
    \includegraphics[width=0.3\textwidth]{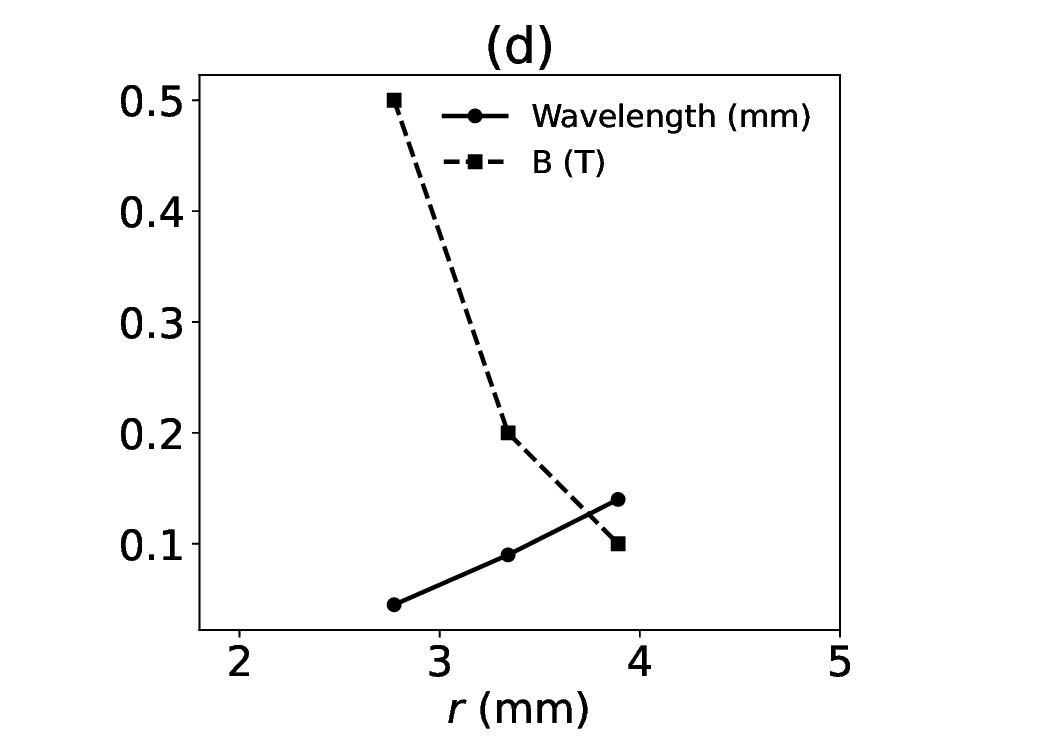}
    \includegraphics[width=0.3\textwidth]{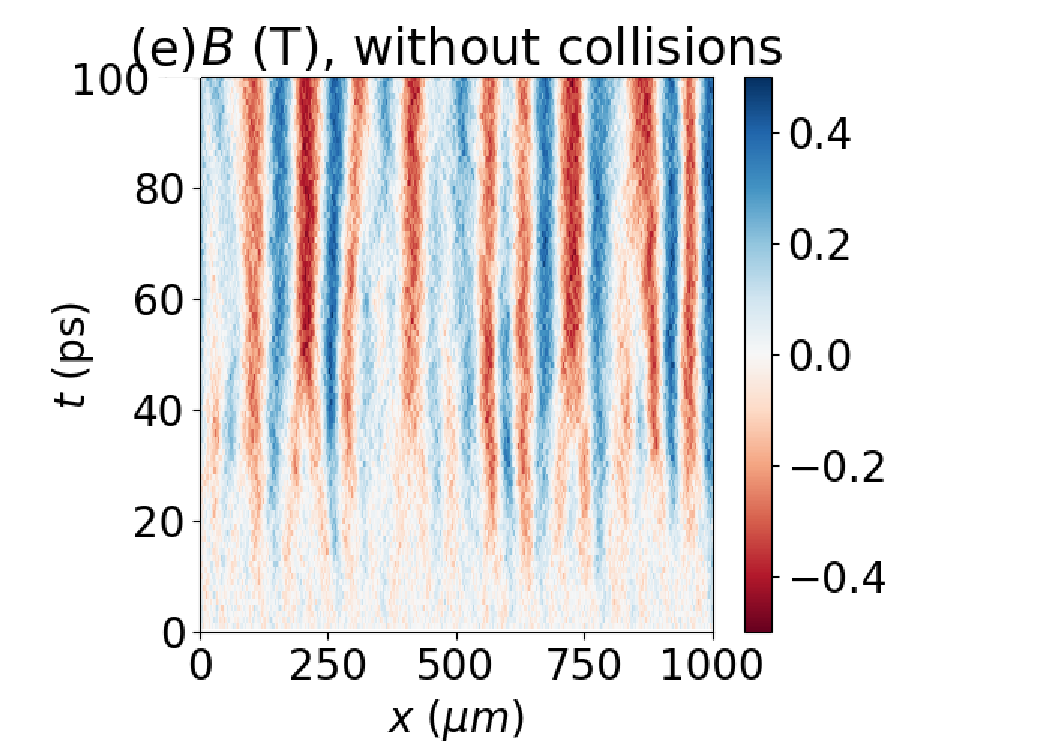}
    \caption{ \label{fig:thxplmj}
   (a) Electron pressure anisotropy versus distance as predicted by our model for the LMJ experiment. The density profiles (red lines) and the locations of the three 1D PIC simulations (markers) are shown on the right axis. (b) Temporal evolution of the $B$-field energy within the PIC simulation domain, comparing cases with (solid line) and without (dashed line) Coulomb collisions. The collisionless prediction from Eq.~\eqref{eq:gnc} is superimposed as a dotted line. (c) Temporal evolution of the electron pressure anisotropy extracted from collisional (solid line) and collisionless (dashed line) simulations. The dotted line indicates an exponential decay at rate $\nu_{ei}$ (d,e) Space-time evolution of the $B$-field for the (d) collisional and (e) collisionless simulations.}
\end{figure*}

\subsubsection{Theoretical analysis}

We performed hydro-rad \textsc{troll} simulations of the experiment for the two target materials. These simulations predict an exponentially decreasing plasma density with a scale length of $L \simeq 400\,\rm \mu m$ for Ti and $L \simeq 290\,\rm \mu m$  for Au, and an electron temperature of $\sim 1\,\rm keV$ in both cases. From these values, we estimate $n_e^\star \simeq 3.3 \times 10^{19}\,\rm cm^{-3}$ for Ti and $n_e^\star \simeq 0.92\times 10^{19}\,\rm cm^{-3}$ for Au. Moreover, we set $r^\star = L$, which is consistent with the transverse extent of the plasma profile.

Under these conditions, the model predictions are shown in Fig.~\ref{fig:thxplmj}(a). In the Ti foil, the electron pressure anisotropy [Fig.~\ref{fig:thxplmj}(a), black curves, left axis] peaks between $2.5 \lesssim r \lesssim 4\,\rm mm$, reaching $a \simeq 0.2$--$0.4$. In the more collisional Au foil, the anisotropy is weaker and more localized, peaking at $a \simeq 0.05$. The density profiles used to derive the anisotropy are indicated by red curves [Fig.~\ref{fig:thxplmj}(a), right axis]. These results demonstrate that the anisotropy develops preferentially in the Ti plasma within a density range of $10^{17} \lesssim n_e \lesssim 10^{19}\,\rm cm^{-3}$.

A preliminary estimate of the dominant Weibel mode can be obtained using the collisionless formulas \eqref{eq:gnc} and~\eqref{eq:knc}. Due to its larger anisotropy, the Ti plasma is more susceptible to the collisionless Weibel instability than the Au plasma. Over a $\sim 100\,\rm ps$ hydrodynamic timescale, we expect the peak growth rate in Ti to yield a seed-field amplification of $\sim 20$ $e$-foldings, which should be sufficient to reach saturation. In the collisionless limit, the dominant wavelength in the Ti plasma ranges from $\sim 50$ to $\sim 200\,\rm \mu m$ in regions of significant anisotropy.

These predictions, however, should be interpreted with caution, as they neglect collisional effects on the Weibel instability, which are potentially significant in high-$Z$ materials. Here, unlike in previous sections, the predicted magnetic wavelength is comparable with, or larger than, the electron mean free path. It is also substantially shorter than the values suggested by the proton radiographs in Fig.~\ref{fig:lmj_p+}. The predicted wavelength varies by a factor of $\sim 4$ within the measurement region, and the radiographs are expected to be smoothed by proton scattering within the Ti target over a $\sim 80\,\rm \mu m$ scale. This smoothing effect complicates precise quantitative analysis.

To address the impact of Coulomb collisions on the Weibel instability, we ran several 1D PIC simulations using the setup described in Sec.~\ref{sec:weibel}. We considered a homogeneous, fully ionized Ti plasma at three different densities, $n_e \in \{1/16, 1/4, 1\} \times 10^{19}\,\rm cm^{-3}$. These values correspond to the locations $r \in \{2.77, 3.34, 3.89\}\,\rm mm$, respectively, marked by red triangles in Fig.~\ref{fig:thxplmj}(a). In all three cases, the electron distribution was initialized with $T_x=T_z = 0.75\,\rm keV$ and $T_y = 1\,\rm keV$. 

For $n_e= 0.25\times 10^{19}\,\rm cm^{-3}$, the integrated $B$-field energy [Fig.~\ref{fig:thxplmj}(b)] shows exponential growth in the collisionless case (dashed line), in agreement with the theoretical value [Eq.~\eqref{eq:gnc}, dotted line]. When including Coulomb collisions (solid line), the $B$-field energy exhibits a shorter, weaker-than-exponential growth phase, and saturates about an order of magnitude below the collisionless case. The $x$--$t$ evolution of the $B$ field [Figs.~\ref{fig:thxplmj}(d,e)], reveals clear modulations after $\sim 5$--$10\,\rm ps$. These are characterized by a similar $\lambda \sim 90\,\rm \mu m$ typical wavelength in both collisional [Fig.~\ref{fig:thxplmj}(d)] and collisionless [Fig.~\ref{fig:thxplmj}(e)] cases, comparable to the value $\lambda_{\rm W} \sim 70\,\rm \mu m$ given by Eq.~\eqref{eq:knc}.

The thermal anisotropy deduced from the spatially averaged electron temperatures [Fig.~\ref{fig:thxplmj}(c)] shows almost no evolution in the collisionless case (dashed line), as expected for this reduced geometry \cite{Bret_Gremillet_2010}. By contrast, with Coulomb collisions included, the solid black line shows a decrease in anisotropy consistent with the equilibration expected from electron-ion momentum exchange [$a \propto \exp(-\nu_{ei}t)$, dotted line]. This suggests that the instability saturates [see solid line in Fig.~\ref{fig:thxplmj}(b)] due to collisional isotropization rather than particle trapping \cite{Davidson_1972}. This regime, discussed in Ref.~\cite{Wallace_87}, results in the sparse patches of decaying magnetic fluctuations observed in Fig.~\ref{fig:thxplmj}(d) after $\sim 40\,\rm ps$.

Clearly, the mechanism responsible for the anisotropy modeled in Sec.~\ref{sec:th}--the quasi-spherical plasma expansion--is not captured by these periodic 1D simulations. In this simplified setup, the system starts anisotropic and naturally relaxes to equilibrium. The saturated field appears to be weaker ($B \simeq 0.1$--$0.2\,\rm T$) with collisions than without ($B \simeq 0.4\,\rm T$). The collisionless formula, Eq.~\eqref{eq:pok}, yields $B_{\rm sat} \simeq 0.5\,\rm T$, in good agreement with the PIC results. Unfortunately, to our knowledge, no collisional counterpart for this estimate currently exists.

Because the Coulomb collision frequency is comparable to the Weibel growth rate, the field growth ceases after a few collisional times. Thus, the maximum $B$-field strength achieved in our collisional simulations is directly linked to the initial seed level. In a realistic system, the competition between collisions and the transverse cooling from spherical expansion should sustain the anisotropy over hydrodynamic timescales, allowing the system to reach Weibel saturation. Consequently, while our numerical results confirm Weibel growth, the saturated $B$-field strength in the experiment cannot be inferred from these simulations or our analytical estimates.

\section{Conclusions and perspectives}
\label{sec:ccl}

During spherical plasma expansion, the transverse electron pressure declines due to the conservation of transverse momentum, leading to a rise in electron pressure anisotropy. This anisotropy is mitigated by electron-ion Coulomb collisions, which tend to restore the equilibrium electron distribution. If collisionality is sufficiently low, the Weibel instability is triggered, generating magnetic filamentary structures. Accurate modeling of this process requires a three-dimensional kinetic treatment that resolves Coulomb collisions. However, such simulations--whether based on Fokker-Planck or PIC methods--remain prohibitively expensive over hydrodynamic spatiotemporal scales unless a reduced spherical geometry is adopted. In this simplified configuration, the quasi-analytical model proposed by True~\cite{True_85} provides an estimate of the anisotropy growth using a closure approximation for the electron distribution function.

The weak electron anisotropy expected in the high-energy laser experiments studied here contrasts with the conditions typically assumed in previous studies on Weibel-unstable laser-driven plasmas~\cite{Schoeffler_2014}. In the collisionless regime, our PIC simulations demonstrate that instability saturation is better captured by a quasilinear treatment~\cite{Pokhotelov_2011} than by the standard trapping-based criterion~\cite{Davidson_1972}. Consequently, by combining the models of Refs.~\cite{True_85, Pokhotelov_2011}, it is possible to estimate the properties of the self-induced magnetic filaments, provided that the spatial density profile and sphere radius are known, electron-electron collisions are negligible (or subdominant), and the electron mean free path exceeds the dominant filament wavelength. These parameters can be inferred from hydrodynamic simulations (Sec.~\ref{sec:peml}) or Thomson scattering measurements (Sec.~\ref{sec:sutcliffe}). Although the assumption of spherical expansion is a strong simplification for planar foil geometries, the main physics of the model should operate at sufficient distances from the foil.

The predicted characteristic wavelength and localization of the instability match with proton radiographs from two independent OMEGA experiments \cite{Masson_2019, Sutcliffe_2022} involving weakly collisional expanding CH plasmas, although the model underestimates the saturated field strength. The neglect of multidimensional effects and extended spatiotemporal scales likely contribute to the observed deviations.

The same mechanism likely accounts for the filamentary structures observed during the ablation of a Ti foil at the LMJ-PETAL facility. These results highlight the critical role of Coulomb collisions: when a higher-$Z$ material such as Au is ablated under identical irradiation conditions, no filamentary structures are observed. In this case, collisions dominate, leading to negligible electron anisotropy and suppressing the Weibel instability.

In an intermediate regime where the collision frequency is comparable to the Weibel growth rate--as observed in Ti plasmas--, no theoretical predictions currently exist for the saturated Weibel field. While PIC simulations in simplified 1D periodic geometry capture the initial growth of the magnetic field, they fail to account for the sustained electron anisotropy driven by spherical expansion. Consequently, these simulations identify the onset of field growth but cannot predict the experimentally relevant saturated fields. Further investigations are necessary to elucidate the mechanisms governing Weibel saturation under such collisional conditions, e.g. following the approach outlined in Ref.~\cite{Bell_2024}.

For the three experiments analyzed, we have demonstrated that the Weibel instability can arise in the expanding corona, generating sub-tesla $B$-fields with $\lambda \sim 100$--$300\,\rm \mu m$ wavelengths in low-density regions. These fields are accessible to characterization via proton radiography; however, their potential impact on hohlraum dynamics--or, more broadly, on high-energy-density experiments--remains an open question. 

\charles{Interestingly, a recently published article examines how the  electron Weibel structures is driven in  a planar expanding plasma \cite{Lezhnin_2026}. The obtained filament orientation and magnetic field strength are not compatible with the experimental observations presented here.}

The magnetic pressure exerted on the self-generated filaments \charles{by the mechanism discussed here} is found to be orders of magnitude below the electron thermal pressure, precluding any significant magnetic pinching effect. Consequently, we do no expect the Weibel instability to drive density perturbations large enough to influence laser propagation through refraction. Yet, despite the relatively weak fields induced, one should consider the possibility of Faraday rotation across extended plasma volumes. For the measured parameters ($\int B\,dl \lesssim 0.1\,\rm T\, mm$ and $n_e/n_c \lesssim 10^{-3}$), the Faraday rotation angle is estimated to be $\phi_{\rm rot} \lesssim 10^{-4}\,\rm rad$ (assuming a laser wavelength of $0.35\,\rm\mu m$). This negligible rotation suggests that Weibel-generated fields are unlikely to affect polarization-dependent instabilities such as cross-beam energy transfer~\cite{PRL_Michel_2009}.

The predicted field strengths and wavelengths also imply weak magnetization over large plasma volumes. While one might speculate about their influence on heat flux models~\cite{POP_Michel_2023} critical to ICF or high-energy-density-physics, the low-density regime ($n_e/n_c \lesssim 10^{-3}$) where thermal anisotropy persists suggests minimal hydrodynamic impact. Thus, Weibel-driven fields should have negligible influence on experimental hydrodynamics under these conditions.

We conclude that this $B$-field generation mechanism primarily affects proton radiographs of expanding plasmas, creating filamentary structures as observed in the experiments detailed here, and poses no apparent threat to ICF-relevant experiments--unless future evidence suggests otherwise. Nonetheless, these experiments represent valuable test cases for hydrodynamic schemes incorporating full-pressure-tensor models with high-order fluid closures, such as those proposed in Ref.~\cite{Sarrat_2016}.

\appendix* 

\section{Derivation of Eq.~\eqref{eq:phi}}
\label{app:ne}

The radial profile of the electron density is obtained by integrating Eq.~\eqref{eq:fe} over the energy $E = m_e v^2/2 - e\phi$ and the angular momentum $l = m_e v_\perp r \equiv \sqrt{u}$:
\begin{align}
    n_e = \frac{1}{r^2} \int_{-e\phi(r)}^\infty dE \int_0^{l_{\rm max}^2} du\,\frac{f_e(E,\sqrt{u})}{v_r(E,\sqrt{u})}  \,,
\end{align}
where $v_r(E,\sqrt{u}) = \sqrt{2/m_e}\sqrt{E+e\phi - u/(2m_e r^2)}$ and $l_{\rm max}^2 = 2m_e r^2(E+e\phi)$. The integration is split into two terms corresponding to the intervals $u \in [0, l_\star^2]$ and $u \in [l_\star^2, l_{\rm max}^2]$. The first integral can be performed analytically:
\begin{align} 
   &\frac{F_0}{r^2} \int_{-e\phi(r)}^\infty dE \int_0^{l_\star ^2} du\, \frac{e^{-E/T_e}}{v_r(E, \sqrt{u})} \nonumber \\ 
   &= n_e^\star \left[e^{\frac{e\phi}{T_e}} - \left(1-\frac{r^2}{r_\star^2}\right)^{1/2} e^{\frac{e\phi}{T_e(1-r^2/r_\star^2)}} \right] \,.
\end{align} 
The second one, which was not detailed in Ref.~\cite{True_85}, is given by
\begin{equation} 
     \bar{n}_e(\phi,r) = \frac{F_0}{r^2} \int_{-e\phi(r)}^\infty dE\,\int_{l_\star ^2}^{l_{\rm max}^2} du\,\frac{e^{-E/T_e - (\sqrt{u}-l_\star )/l_0}}{v_r(E,\sqrt{u})} \,,
\end{equation}
and requires a numerical evaluation. For the plasma parameters considered here, replacing $e^{-(\sqrt{u}-l_\star)/l_0}$ with $e^{-(l_{\rm max}-l_\star)/l_0}$ provides a robust approximation. This simplifies the numerical evaluation to
\begin{align} \label{eq:nebar_final}
     \bar{n}_e(\phi,r) &\simeq  F_0\,(2m_e)^{3/2}\int_{-e\phi}^\infty dE\, e^{-E/T_e - [l_{\rm max}(E)-l_\star(E) ]/l_0} \nonumber \\
     &\times \sqrt{E+e\phi - l^2/(2m_er^2)} \, .
\end{align}
We have employed the above expression in combination with Eq.~\eqref{eq:phi} to numerically determine $\phi$ as a function of the radial position $r$.

\section*{Acknowledgements}

This work has been done under the auspices of CEA-DAM and
the simulations were performed using HPC resources at TGCC/CCRT and CEA-DAM/TERA.
We acknowledge M.~Manuel for valuable input and exchange, and D.~Bénisti for insightful comments. 

\section*{Data availability}

The data that support the findings of this study are available from the corresponding authors upon reasonable request.

%
\end{document}